# Metallic and All-Dielectric Metasurfaces Sustaining Displacement-Mediated Bound States in the Continuum

*Luca M. Berger [1], Martin Barkey [1], Stefan A. Maier [2,3,1], Andreas Tittl [1,\*]*

[1] *Chair in Hybrid Nanosystems, Nanoinstitute Munich, and Center for NanoScience, Faculty of Physics, Ludwig-Maximilians-University Munich, Königinstraße 10, 80539 München, Germany*
[2] *School of Physics and Astronomy, Monash University, Wellington Rd, Clayton VIC 3800, Australia*
[3] *Department of Physics, Imperial College London, SW7 2AZ London, United Kingdom*
E-mail: Andreas.Tittl@physik.uni-muenchen.de

**Abstract**

Bound states in the continuum (BICs) are localized electromagnetic modes within the continuous spectrum of radiating waves. Due to their infinite lifetimes without radiation losses, BICs are driving research directions in lasing, non-linear optical processes, and sensing. However, conventional methods for converting BICs into leaky resonances, or quasi-BICs, with high-quality factors typically rely on breaking the in-plane inversion symmetry of the metasurface and often result in resonances that are strongly dependent on the angle of the incident light, making them unsuitable for many practical applications. Here, we numerically analyze and experimentally demonstrate an emerging class of BIC-driven metasurfaces, where the coupling to the far field is controlled by the displacement of individual resonators. In particular, we investigate both all-dielectric and metallic as well as positive and inverse displacement-mediated metasurfaces sustaining angular-robust quasi-BICs in the mid-infrared spectral region. We explore their behavior with changes in the incidence angle of illumination and experimentally show their superior performance compared to two conventional alternatives: silicon-based tilted ellipses and cylindrical nanoholes in gold. We anticipate our findings to open exciting perspectives for bio-sensing, conformal optical devices, and photonic devices using focused light.

**1. Introduction**

The development of optical metasurfaces to tailor light-matter interactions on the nanoscale has led to breakthroughs from negative refraction,[1] energy conversion,[2] and ultrathin optical elements[3] to photonic computation.[4] In the quest for improved and diversified metasurface functionality, features like high-quality factors (Q factors, defined as the quotient of resonance frequency and linewidth[5]), long resonance lifetimes, and high electric field confinement have been explored in the literature.[6] To reach these goals, bound states in the continuum (BICs), first described in quantum physics[7,8] and later found to be a phenomenon applying to waves in general, have emerged as a pioneering technology. BICs are discrete and spatially bounded yet exist within the energy range of continuous states. Mathematically, they are vortex centers in the polarization directions of far-field radiation.[9]

True BICs have been theoretically predicted in infinite periodic arrays, by appropriately tuning the parameters in the wave equation, either exploiting its separability due to symmetry, tuning its system parameters to find accidental BICs, or by reverse engineering them from an artificial potential.[10] Moreover, BICs can exist in single isolated structures as supercavity modes.[11,12] In general, by detuning the system parameters from the BIC condition, a leaky channel is created and the resonance can couple to the continuum with a finite Q factor, called quasi-BIC (q-BIC).[10] Various configurations of nanostructures have been explored to yield high-Q factors in the literature, such as one[13,14] and two-dimensional[15] periodic arrays or radial arrangements.[16]

Typically, metasurfaces based on all-dielectric resonators arranged into a periodic array break the in-plane inversion symmetry to offer both high-Q factors and strong near-field enhancement, enabling state-of-the-art technologies for sensing[15,17] and enhanced light-matter interactions.[18] A similar effect emerges for symmetric metasurfaces sustaining BICs at normal incidence when the angle of incidence is tuned.[19] Another desired aspect of metasurfaces is control over the ratio of their radiative to intrinsic loss rates $\gamma_e/\gamma_i$ that allows, for instance, for transitions between electromagnetically induced transparency and absorption.[20]

Apart from a few exceptions,[21,22] studies on nanophotonic or plasmonic metasurfaces yielding high Q-resonances have exclusively focused on (near) normal incidence or angle-multiplexed illumination with experiments focusing on refractive microscope objectives.[23] Unsurprisingly, most conventional metasurfaces based on q-BICs or other nanophotonic or plasmonic principles feature resonances that shift strongly with the angle.[23] While this feature has been exploited for angle-multiplexed sensing,[23] these metasurfaces cannot be used for applications requiring angular robustness (i.e., an optical response that does not vary with incidence angle), such as conformal optical devices, flexible substrate technology, displays, and photonic devices using focused light.[21] To date, there are only very few mentions of angular robust metasurfaces in the literature[21,24,25] with a further paucity of works aimed at a theoretical understanding of angular robustness.[26,27] For example, one study proposed a gold-based metasurface featuring a resonance that can be excited at a small range of angles wherein it shows angular robustness.[19] However, the proposed metasurface requires femtosecond laser writing in photoresist to make structures with different heights which is difficult with electron beam lithography. A fundamental and systematic study on angular-robust BIC-driven metasurfaces is still missing.

Refractive objectives require multiple glass elements with different Abbe numbers and refractive indices to control chromatic aberrations.[28] This strongly limits the wavelength range in which refractive objectives can operate and makes them unsuitable for broadband sensing or white light applications.[28] In contrast, reflective microscope objectives use mirrored surfaces to focus light, a process inherently independent of wavelength within a larger spectral range, with the advantage of high numerical aperture, large working distance, and no chromatic aberration.[29–31] Therefore, reflective optical systems are the objectives-of-choice when broadband achromatic focusing is required, especially in the NIR to MIR range.[32] The main drawback of reflective objectives in connection with optical metasurfaces is the difficulty of realizing angular-robustness within the operational range of polar and azimuthal angles.[21]

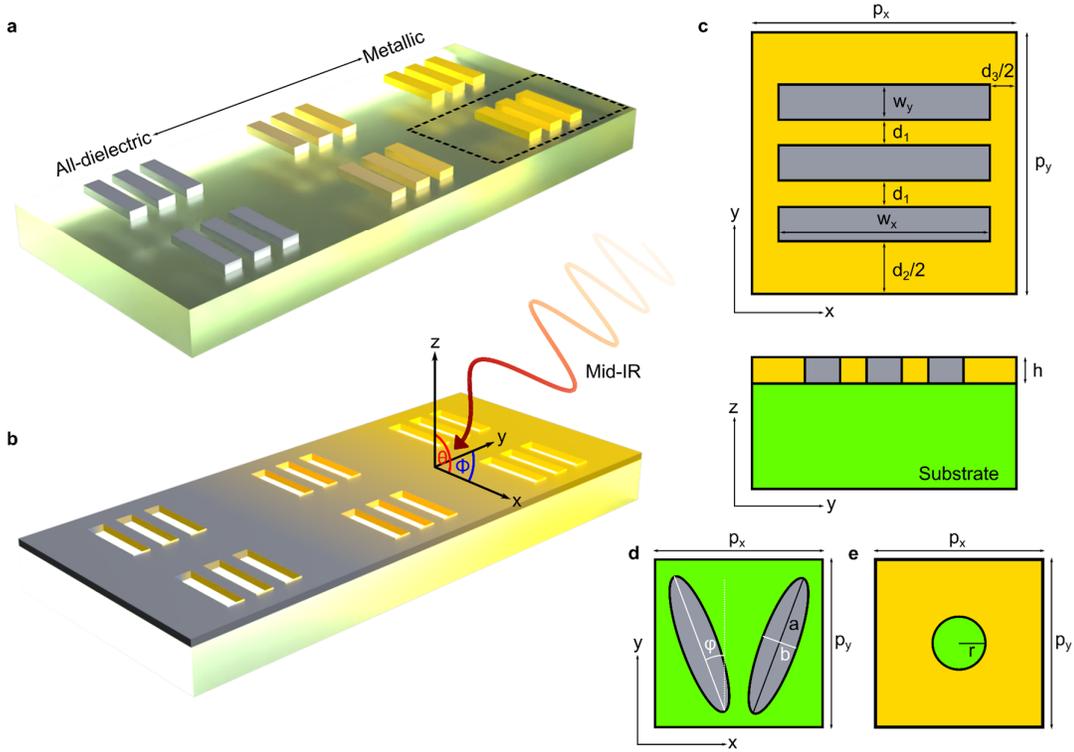

**Figure 1.** The displacement-mediated quasi-BIC metasurface. (a-b) Schematic of the superlattice metasurface made of periodic arrangements of all-dielectric and metallic (a) positive and (b) inverse beams. (c-d) The unit cell of the displacement-mediated q-BIC metasurface from the (c) top and side view. The three beams making up the superlattice have the dimensions $w_x \times w_y \times h$. They are separated in $y$ from each other by a distance $d_1$ and to the next set of three beams by a distance $d_2$ and in $x$ by a distance $d_3$. (d) The unit cell of the tilted ellipses metasurface. The ellipses with a height $h$ are defined by a long axis $a$ and short axis $b$. The BIC is converted into a leaky resonance by breaking the in-plane inversion symmetry by tilting the ellipses under an angle $\varphi$ with respect to the $y$-axis giving an asymmetry $\alpha_e = \sin(\varphi)$. (e) The unit cell of the gold nanohole metasurface consisting of cylindrical holes with a radius $r$ and a height $h$. The dimensions of all unit cells are defined by $p_x \times p_y$.

Here, to the best of our knowledge, we experimentally realize for the first time a metasurface that can convert a BIC into a guided-mode resonance through relative displacement tuning while maintaining the in-plane inversion symmetry. We demonstrate the tolerance of this resonance toward changes in the angle of incident light showing strong performance in combination with a high numerical aperture (NA) reflective objective. We investigate both metallic and all-dielectric, positive (**Figure 1a**) and inverse (**Figure 1b**), displacement-mediated quasi-BIC metasurfaces. By inverting the structure, we mean that the material-filled parts of the positive structure become air (or other surrounding material) while the parts that were air (or surrounding material) become structural material. We confirm Babinet's principle, which allows for a straightforward implementation of an inverse (positive) structure if the positive (inverse) one is already understood, thereby broadening its range of applications. Babinet's principle states that, in the limit of thin metallic films with high conductivity, if a structure features a resonance in transmission (reflection) its inverse structure will feature a similar resonance in reflection (transmission) under a 90-degree change in polarization, respectively, as long as their geometrical parameters are the same.[33] For example, inverse-rod metallic connected films have recently shown improved in situ characterization of electrochemical interfaces.[34] This would not have worked with rods as a connected metallic film was ubiquitous to function as the working electrode.

Unlike conventional q-BIC-based platforms, our metasurface does not break the in-plane inversion symmetry. Instead, the BIC is converted into a leaky resonance by detuning the displacement of three equally spaced beams ($w_x \times w_y \times h$) forming the unit cell (**Figure 1c**) to adjacent unit cells in $y$. A displacement asymmetry parameter $\alpha$ can be defined as

$$\alpha = \frac{p_y}{p_{y0}} - 1 \qquad (1)$$

$$p_{y0} = 3d_1 + 3w_y \qquad (2)$$

where $p_{y0}$ is the unit cell length in $y$ ($p_y$) when the distance between the beams within the unit cell $d_1$ and the distance between two beams of adjacent unit cells $d_2$ are the same. Since the beams are made of silicon or gold in air or air in silicon or gold (Figure 1c) our metasurface can be regarded as a superlattice.[35]

Three superlattice metasurfaces, one silicon-based and two gold-based, are experimentally compared to two conventional metasurfaces used in the mid-IR spectral region, the Si-based tilted ellipses (**Figure 1d**) and gold nanohole (**Figure 1e**) metasurfaces. Both periodically arranged all-dielectric tilted ellipses and metal nanohole metasurfaces have led to cutting-edge breakthroughs in biospectroscopy,[17,23,36,37] catalysis,[5,38] higher harmonic photon generation,[39,40] and more exotic light-matter effects[41,42]. Consequently, they serve as a valuable platform for comparison. We numerically investigate the angular behavior of our metasurfaces and confirm the correct operation of our metasurfaces under near-normal incidence illumination. Then, the feature of angular robustness is explored by illuminating the samples under a reflective microscope objective. We complete our studies by performing a detailed comparison of performance metrics between the metasurfaces, with special attention to the relative changes in the resonances as the range of incidence angles is increased.

## 2. Results and discussion
### 2.1. Numerical design of the superlattice

The unit cell of our displacement-mediated q-BIC metasurfaces (Figure 1c) forms the basis for the numerical simulations. When $\alpha$ is varied for the positive silicon superlattice (**Figure 2a**) a BIC above the Rayleigh anomaly (RA) is converted into a leaky resonance when the incident light is polarized with the electric field parallel to the long axis of the beams (**Figure 2b**). The RA is a phenomenon associated with light diffracted parallel to the surface of a periodic structure.[43] When a resonance occurs on the spectrally higher frequency side of the RA the resonance lifetime and electric near-field enhancement are strongly reduced.[44] Therefore, we will limit our investigations to the modes appearing on the spectrally longer wavelength side of the RA. Since the RA is proportional to the unit cell size

$$\Delta\alpha \sim \Delta\lambda_{RA}, \tag{3}$$

where $\lambda_{RA}$ is the center wavelength of the RA. For $\alpha<0.3$, the q-BIC resonance is spectrally significantly separated from the RA and yields strongly modulating sharp resonances (**Figure 2c**). The numerical multipolar expansion of the positive silicon superlattice performed by Shi *et al*.[35] confirmed the nature of the q-BIC and revealed the primary contributing modes as the toroidal dipole and magnetic quadrupole modes. At $\alpha=0.1$ a maximum local electric field enhancement $|E/E_0|_{max}$ of 39 is obtained, where $E_0$ and $E$ is the incident and local electric field at $z=h/2$ (**Figure 2d**). This value shares an inverse relationship with α. Electric charge accumulates in an alternating fashion around the short edges of the beams.

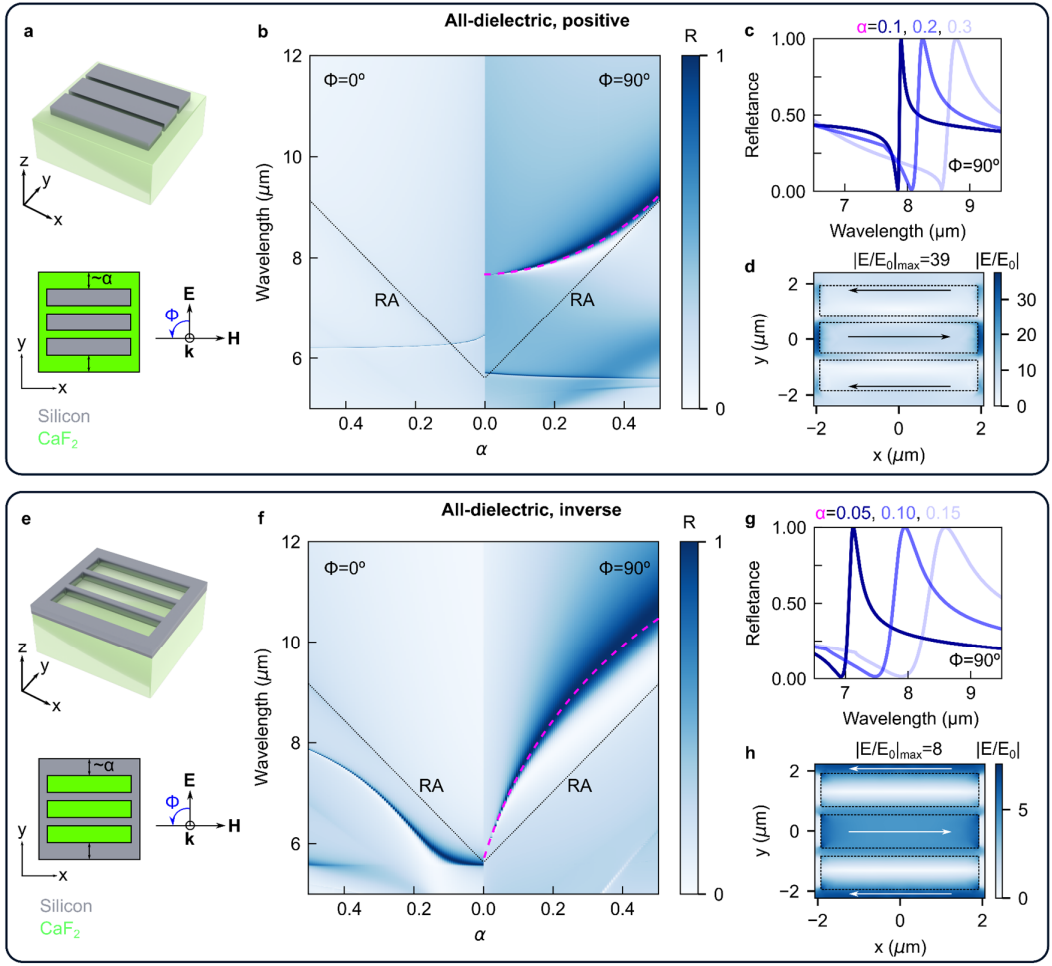

**Figure 2.** Numerical design of the all-dielectric (Si) displacement-mediated q-BIC metasurface. (a) Schematics of the unit cell of the positive silicon superlattice including the polarization of the normal incident light considered in the simulations, where E and H are the incident electric and magnetic field vectors, and k is the Poynting vector. $\Phi$ rotates the field vectors around the k axis. (b) The numerical reflectance spectra of the positive silicon superlattice, with $w_x$=3.85 µm, $w_y$=1.1 µm, $d_1$=$d_3$=275 nm, $h$=650 nm, recorded while varying $\alpha$ from 0 to 0.5 when $\Phi$ is 0° (left half) and 90° (right half). At $\Phi$=90° the q-BIC (pink-dashed line) appears around 8 µm and spectrally shifts closer to the RA with increasing $\alpha$. (c) Selected reflectance spectra showing the q-BIC resonance at $\alpha$=0.1, 0.2, 0.3. (d) Electric near-field distribution $|E/E_0|$ for $\alpha$=0.1 at $z$=$h/2$, including schematic electric field vectors (black arrows), with a local maximum of 39. (e) Schematics of the unit cell of the inverse silicon superlattice similarly to (a). (f) The numerical reflectance spectra of the inverse silicon-based superlattice with the parameter definitions provided in (b) while varying $\alpha$ from 0 to 0.5 when $\Phi$ is 0° (left half) and 90° (right half). At $\Phi$=90° the q-BIC (pink-dashed line) appears on the longer wavelength side of the RA. (g) Selected reflectance spectra showing the q-BIC resonance at $\alpha$=0.1, 0.2, 0.3. (h) Electric near-field distribution $|E/E_0|$ for $\alpha$=0.05 at $z$=$h/2$, including schematic electric field vectors (white arrows), with a local maximum of 8.

The mechanism of q-BIC formation can be understood from the electric near-field plots. At the BIC condition, each adjacent beam has the local electric field confined to its core polarized along its long axis but in opposing directions to impede far-field radiation via cancellation of dipole moments. When a collection of beams is periodically displaced, the electric field in the central beam remains the same while the electric field of the outer two beams is pressed towards the outer edges of the beams, where the field scatters into the far field (Figure 2d).

We find that for the superlattice made of silicon, there is an inverse structure that works similarly to its positive counterpart. However, to the best of our knowledge, there is no analytical theory linking the physical properties of two inverse all-dielectric structures. The unit cell of the inverse silicon-based superlattice with the same parameters as its positive counterpart (**Figure 2e**), similarly to it, produces a q-BIC resonance for $\alpha \neq 0.1$ for light polarized with its electric field aligned parallel to the long axis of the beams (**Figure 2f**). Both the positive and inverse all-dielectric superlattice feature a q-BIC resonance above the RA under the same polarization.

Interestingly, as $\alpha$ is increased, the q-BIC resonance shifts to longer wavelengths faster than the RA, which opposes the behavior of its positive counterpart. Since the same definitions apply to the inverse as to the positive structure, the RA shift is proportional to equation (3). Compared to the positive silicon superlattice, the inverse's q-BIC resonance maintains a similar albeit more symmetrical Fano-type line shape (**Figure 2g**). Moreover, its resonance increases in strength and detunes from the BIC condition at a faster rate.

For $\alpha=0.05$ the local electric near field reaches a maximum value of 8 which is considerably lower than for its positive counterpart with a maximum value of 39 at $\alpha=0.1$. The lower $\alpha$ for the inverse structure was taken due to the q-BIC resonance detuning faster from the BIC condition with increasing $\alpha$. Due to the lower field enhancement, electric charge accumulates less around the short edges of the beams, compared to the positive silicon-based superlattice. The electric field vectors appear similar to the positive structure. Due to the significantly higher local electric near field, the positive structure is expected to be the better candidate for sensing applications.

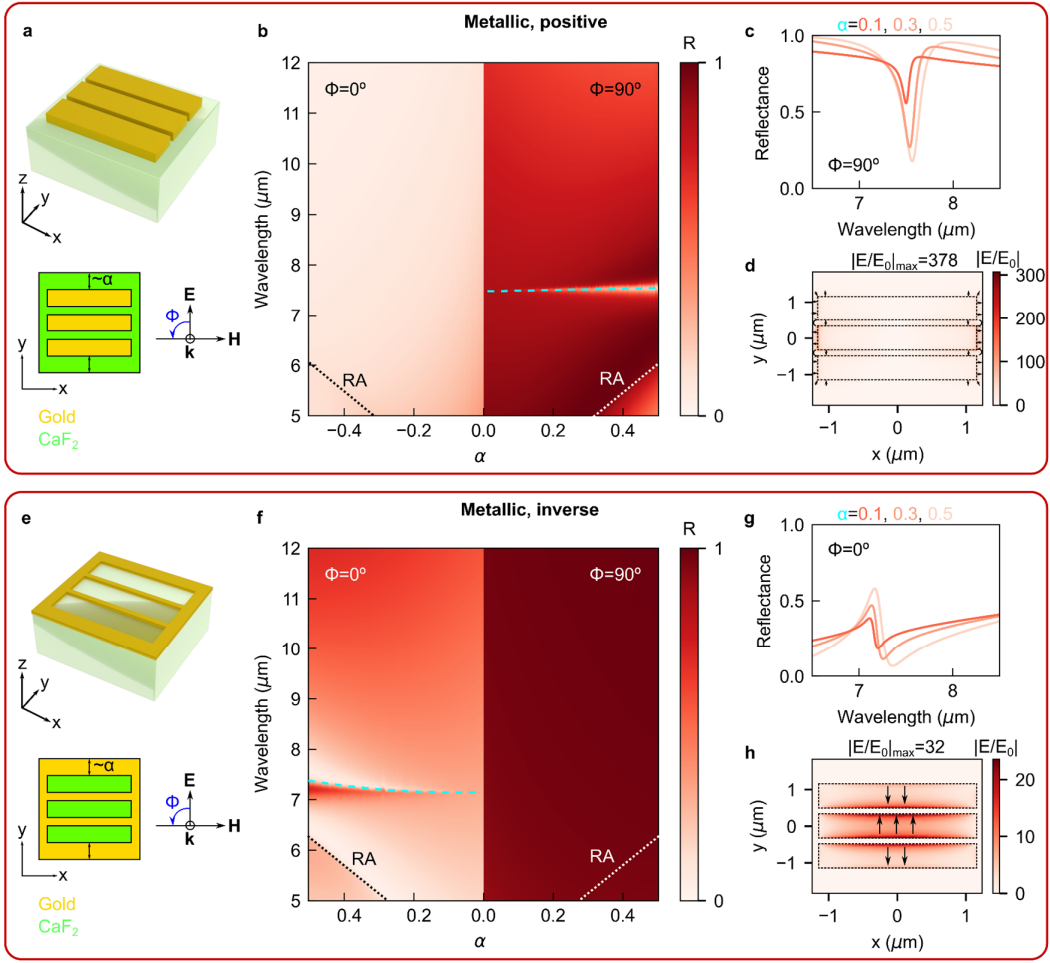

**Figure 3.** Numerical design of the metallic (gold) displacement-mediated q-BIC metasurface. (a) Schematics of the unit cell of the positive gold superlattice including the polarization of the normal incident light considered in the simulations, where E and H are the incident electric and magnetic field vectors, respectively, and k is the Poynting vector. $\Phi$ rotates the field vectors around the k-axis. (b) The numerical reflectance spectra of the positive silicon superlattice, with $w_x$=2.31 µm, $w_y$=660 nm, $d_1$=$d_3$=165 nm, $h$=100 nm, recorded while varying $\alpha$ from 0 to 0.5 when $\Phi$ is 0° (left half) and 90° (right half). At $\Phi$=90° the q-BIC (azure-dashed line) appears around 7.5 µm. (c) Selected reflectance spectra showing the q-BIC resonance at $\alpha$=0.1, 0.3, 0.5. (d) Electric near-field distribution $|E/E_0|$ for $\alpha$=0.3 at the air/CaF$_2$ interface, including schematic electric field vectors (black arrows), with a local maximum of 378. (e) Schematics of the unit cell of the inverse silicon superlattice. (f) The numerical reflectance spectra of the inverse gold-based superlattice with the parameter definitions given in (b) while varying $\alpha$ from 0 to 0.5 when $\Phi$ is 0° (left half) and 90° (right half). At $\Phi$=0° the q-BIC (azure-dashed line) appears on the longer wavelength side of the RA. (g) Selected reflectance spectra showing the q-BIC resonance at $\alpha$=0.1, 0.3, 0.5. (h) Electric near-field distribution $|E/E_0|$ for $\alpha$=0.3 at the air/CaF$_2$ interface, including electric field vectors (black arrows), with a local maximum of 32.

Changing the material of the beams from silicon to gold (**Figure 3a**), the numerical reflectance spectra look qualitatively similar between the positive all-dielectric and metal superlattice (**Figure 3b**). However, there are a few important differences. First, the q-BIC resonance exhibits only small spectral shifts and grows significantly slower with increasing $\alpha$ compared to the positive silicon counterpart. Furthermore, while the RA shifts according to equation (3), the q-BIC resonance is spectrally located significantly further away from the RA, on its higher wavelength side. Since the losses are higher in gold than in silicon, the q-BIC resonances in gold modulate less with lower Q-factors (**Figure 3c**), which is to be expected. However, these results appear very promising due to the symmetric line shape and high maximum local electric near-field enhancement of 378 found for $\alpha=0.3$ (**Figure 3d**), which can be further increased by decreasing $\alpha$.

When $\alpha=0$, the positive and inverse gold superlattices (**Figure 3e**) become simple plasmonic rods and slots, respectively. Compared to previous plasmonic rod antenna geometries with a maximum achievable electric near-field enhancement of around 50,[33] our positive gold superlattice exceeds the obtained $|E/E_0|_{max}$ by at least one order of magnitude. In this case, the BIC condition consists of canceling out magnetic dipoles that suppress the far-field radiation.

The inverse gold superlattice (Figure 3e) features a q-BIC resonance for $\alpha\neq0$ at normal incidence with the electric field vector polarized perpendicular to the long edge of the slots (**Figure 3f**). As Babinet's principle predicts, a nanostructure's line shape in reflection (transmission) should match its inverse's in transmission (reflection) under a 90 degrees change in polarization. For this reason, the line shape of the inverse gold superlattice is less symmetric (**Figure 3g**), as it matches the transmission spectrum of the positive gold superlattice under a 90 degrees change in polarization (see Figure S1 in the Supplementary Information). Consequently, the inverse gold superlattice can be favorable for many applications requiring a symmetric resonance line shape in transmission, such as biosensing[17].

Due to Babinet's principle, the electric near field vector of the inverse gold superlattice (**Figure 3h**) is rotated by 90 degrees aligned parallel to the short axis of the slots in comparison to its positive counterpart (Figure 3d). Here. the BIC condition consists of canceling out electric dipoles. When $\alpha$ is increased the electric or magnetic dipoles no longer perfectly cancel and a scattering channel to the far field is opened. For $\alpha=0.3$ $|E/E_0|_{max}$ is 32 for the inverse gold superlattice, which is of the same order of magnitude as plasmonic slots.

## 2.2 Numerical analysis with a focus on angular behavior

A straightforward approach to predict how robust a structure's resonance is towards changes in the angle of incident light is to track the resonance as the polar angle is varied for a fixed azimuthal angle for *s* (TE) and *p* (TM) polarized light, resulting in an angular dispersion plot (Figure 4 a-e). Only for metasurfaces with a unit cell whose symmetry belongs to the rotational groups $C_n$, for n > 2, there is polarization independence of the structure at normal incidence illumination.[45,46] For non-normal angles of incidence the degeneracy of the modes is typically lifted.

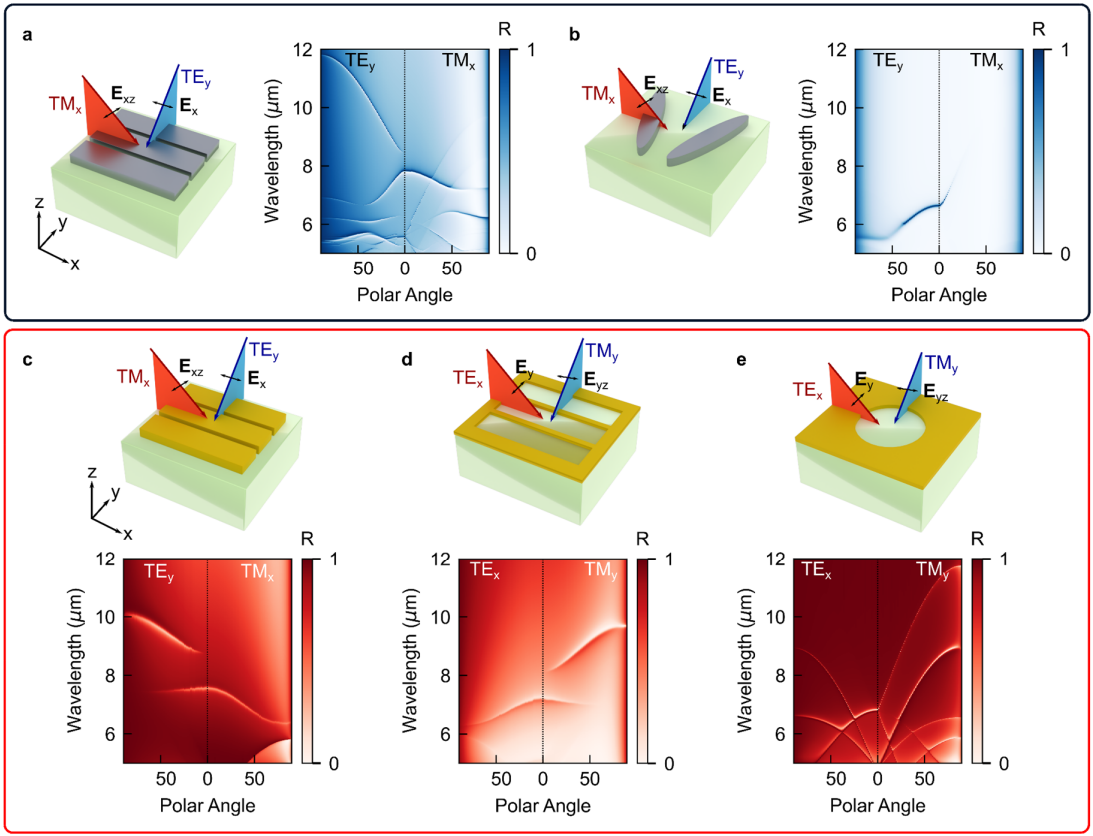

**Figure 4.** Angle sweeps of the displacement-mediated q-BIC and conventional rival metasurfaces. (a-e) Illustrative schematic and corresponding angular dispersion plots (polar angle vs reflectance) for *s* (TE) and *p* (TM) polarized light oriented to excite the primary resonance modes of the silicon (a) positive superlattice and (b) tilted ellipses metasurface, the gold (c) positive and (d) inverse superlattice and (d) nanohole metasurface. The polar angle was varied from 0° to 89°. The structural parameters underlying the polar angle dispersion plots are (a) $\alpha=0.1$, $w_x=3.85$ µm, $w_y=1.1$ µm, $d_1=d_3=275$ nm, $h=650$ nm, (b) $\alpha=\sin(20°)$, $a=3000$ nm, $b=720$ nm, $p_x=4650$ nm, $p_y=3075$ nm, $h=650$ nm, (c-d) $w_x=2.31$ µm, $w_y=660$ nm, $d_1=d_3=165$ nm, $h=100$ nm, and (e) $r=1$ µm, $p_x=p_y=5$ µm, $h=100$ nm.

The properties extracted from the angular dispersion plots can predict a metasurface's performance under a reflective objective. A reflective objective can be paired with a polarizer that will define the polarization of incident light and will allow for the range of TE to TM polarized light across a window of polar angles.[21] For this reason, two crucial factors affecting the performance of metasurface resonances under reflective objectives are the magnitude and direction of the angular-driven shift in the resonance position for TE and TM polarized light.

The angular dispersion of the main displacement-mediated q-BIC resonance emerging for $\Phi$=90° on the longer wavelength side of the RA was tracked (**Figure 4a**). Interestingly, there is a small resulting shift to shorter wavelengths in the resonance position of the displacement-mediated q-BIC resonance around $\lambda$≈7.8 µm for $TM_x$ incident light for increasing polar angles $\theta$. This shift is similar but smaller compared to silicon tilted ellipses metasurfaces showing a resonance at around $\lambda$≈6.6 µm (**Figure 4b**). All-dielectric tilted ellipses produce a q-BIC mode for *x*-polarized incident light with an asymmetry parameter $\alpha$=sin($\varphi$), where $\varphi$ is the angle that the major axis of the ellipses make with the *y*-axis, also called tilt angle.[47] Almost mirroring across the $\theta$=0° axis in the angular dispersion plot (Figure 4a) the displacement-mediated q-BIC mode shifts to shorter wavelengths at a slightly faster rate with increasing $\theta$ for $TE_y$ incident light until it meets its BIC point at around $\theta$≈50°. In contrast, the q-BIC mode of the silicon tilted ellipses metasurface under $TM_x$ polarized incident light strongly shifts to longer wavelengths until it disappears (Figure 4b). Consequently, the relatively small and directionally similar (shift to shorter wavelengths) angular-driven shift in the $TM_x$ and $TE_y$ displacement-mediated q-BIC resonance position indicates a better performance under reflective objectives for the positive silicon superlattice over the silicon tilted ellipses metasurface. For the silicon superlattice, a second mode emerges for *s* polarized light with its BIC condition at $\theta$=0° (Figure 4a). This mode was numerically described by Shi *et al*.[35] Higher order modes on the shorter wavelength side of the displacement-mediated q-BIC mode appear that are outside the scope of this article.

The angular behavior of the gold superlattice follows the same pattern as the silicon superlattice, although the q-BIC resonance shifts less strongly to shorter wavelengths. The resonance of the positive (**Figure 4c**) and inverse (**Figure 4d**) gold superlattice can be excited for $TE_y$ and $TM_x$ and $TE_x$ and $TM_y$ polarized light, respectively. It gradually disappears with increasing $\theta$. As for the silicon counterpart, the gold superlattice features an emerging q-BIC resonance for $\theta$≠0 for TE$_y$ and TM$_y$ polarized incident light for its positive and inverse variations, respectively. Higher order modes of the gold superlattices are spectrally more separated.

The gold nanohole metasurface has a more complex band structure (**Figure 4e**). It produces a Fano resonance caused by multiple resonance interference at normal incident illumination where the Q-factor increases with decreasing radius *r*.[36] The degeneracy of the primary resonance mode at ca. 6.8 µm at normal incidence is lifted for $\theta\neq0$ for TM polarized light. The upper branch strongly shifts to longer wavelengths due to the RA's similar spectral shift with increased incidence angles according to $\lambda_{RA} = p_{x,y}(n - \sin(\theta_{x,y}))$,[48] where *n* is the refractive index of the substrate and $p_x=p_y$ is the same for the nanohole metasurface. Under TE illumination the nanohole metasurface's resonance shifts to shorter wavelengths. Therefore, both versions of the gold superlattices are expected to show superior performance over the nanohole array when switching from a refractive to a reflective microscope objective or for other applications requiring angular-robust resonances.

## 2.3 Near-normal incidence mid-IR spectral imaging

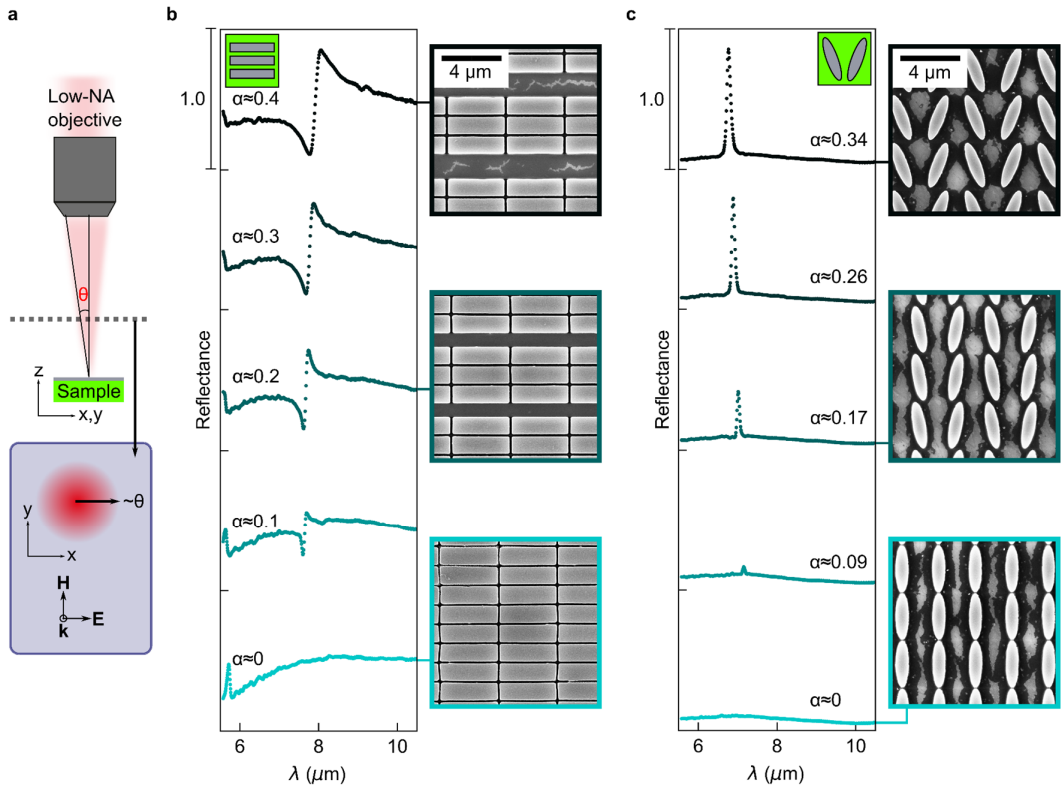

**Figure 5.** Near-normal incidence mid-IR spectral imaging of the all-dielectric superlattice and the tilted ellipses metasurface. (a) Schematic of the objective used and polarization of the incident light focused on the samples of the mid-IR spectral imaging microscope used for the following near-normal incidence optical measurements. (b) The measured reflectance spectra for the positive silicon superlattice for $α≈0$, 0.1, 0.2, 0.3, 0.4 showing the displacement-mediated q-BIC resonance appearing for $α>0$. The adjacent scanning electron microscopy (SEM) images give the structural parameters of the structure as $w_x≈3.7$ µm, $w_y≈1.1$ µm, $d_1≈d_3≈120$ nm, $h≈650$ nm. (c) The measured reflectance spectra for the silicon tilted ellipses metasurface for $α$ varied from 0 to ca. 0.34. The adjacent SEM images give the structural parameters of the structure as $a≈3.3$ µm, $b≈760$ nm, $p_x≈2.41$ µm, $p_y≈3.12$ µm, $h≈650$ nm. Schematics of the relevant unit cell and the polarization of light are provided in the insets (b-c).

Mid-IR spectral imaging with a low numerical aperture (NA≈0.15) focusing objective (**Figure 5a**) of the positive silicon superlattice produced the predicted appearance of the displacement-mediated q-BIC resonance for $α>0$ (**Figure 5b**). The asymmetric line shape was predicted by the simulations (Figure 2c) but can be made more symmetric by appropriate parameter tuning. In contrast, the silicon tilted ellipses metasurface showed a symmetric q-BIC resonance emerging for $α>0$ with a low baseline reflectance of less than ca. 12% (**Figure 5c**). This result was expected as tilted ellipses are a staple q-BIC-based platform-of-choice for biosensing in the literature.[17,23]

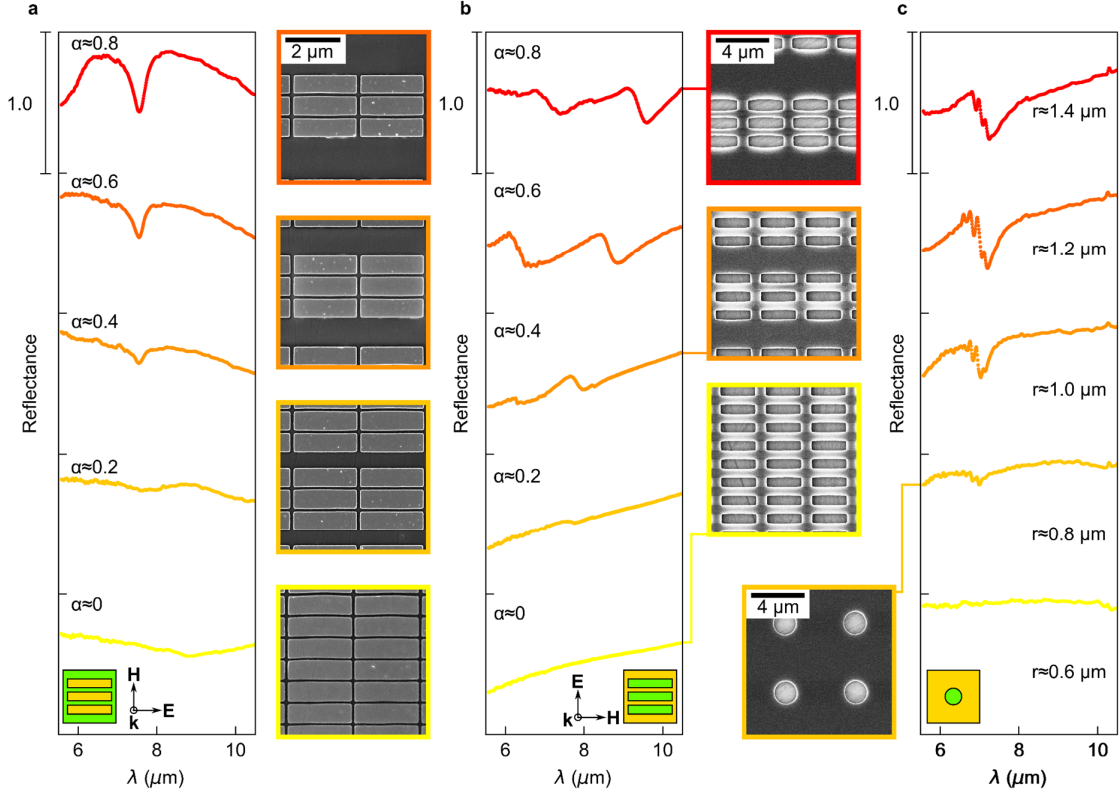

**Figure 6.** Near-normal incidence mid-IR spectral imaging of the all-dielectric superlattice and the tilted ellipses metasurface. (a-b) The measured reflectance spectra for the (a) positive and (b) inverse gold superlattice for $α≈0$, 0.2, 0.4, 0.6, 0.8 showing the displacement-mediated q-BIC resonance appearing for $α>0$. The adjacent SEM images provided the structural parameters of the structure as (a) $w_x≈2.3$ µm, $w_y≈680$ nm, $d_1≈d_3≈100$ nm, $h≈70$ nm, and (b) $w_x≈2.5$ µm, $w_y≈670$ nm, $d_1≈630$ nm, $d_3≈780$ nm, $h≈30$ nm. (c) The measured reflectance spectra and an SEM image for the gold nanohole metasurface for $r≈0.6, 0.8, 1, 1.2, 1.4$ µm, $p_x≈p_y≈5$ µm, $h≈30$ nm. Schematics of the unit cell and the polarization of light are provided in the insets (a-c).

Near-normal incidence mid-IR reflectance spectra of the positive gold superlattice (**Figure 6a**) produced a displacement-mediated q-BIC resonance with a symmetric line profile emerging when $α>0$ as predicted by the numerical models (Figure 3b). As predicted by the simulations (Figure 3b and 3f), the q-BIC resonance of both the positive (Figure 6a) and inverse (**Figure 6b**) gold superlattice grew slower with $α$ compared to the silicon version (Figure 5b). The asymmetric profile of the displacement-mediated q-BIC resonances for the inverse gold superlattice measured experimentally was expected from the simulations (Figure 3g). However, the overall quality of the resonances can be increased by improving the fabrication methods. Since the inverse structures were fabricated with a negative resist which is significantly more sensitive to electron beam exposure and strongly limits the achievable resolution, the distances between the inverse beams or slots had to be made very large, with $d_1 \approx 630$ nm and $d_3 \approx 780$ nm. Due to the larger gaps and unit cell the RA was shifted to longer wavelengths cutting into and shifting the q-BIC resonance (Figure 6b). Moreover, negative resists suffer from a negatively impacting overcut profile that limited the inverse gold superlattice's metal film thickness to maximally 30 nm. In contrast, its positive complementary structure was made with a thickness of 70 nm, which allowed for stronger resonances.

The two gold superlattices produced cleaner line profiles compared with a gold nanohole metasurface (**Figure 6c**). The latter showed extra small modulations of the main resonance appearing around ca. 7 µm, which are explained by the numerically modeled angular dispersion plots (Figure 4e). Objectives with a non-zero NA take in light from a range of angles. The measured spectrum is formed by the average light power transmitted by the objective, which depends on the range of angles with which it can accept incoming light. Therefore, the measured spectrum can be predicted by averaging the spectra corresponding to the range of admissible angles and weighing them according to their relative contribution or power with which they pass the objective. Even for small NA refractive objectives (Figure 6a), the measured spectrum is affected by the relative contributions it receives from light arriving across a small range of angles. The small modulations in the spectrum close to the resonance of the gold nanohole metasurface were caused by its main resonance at normal incidence splitting for small angular changes (Figure 6c).

**2.4 Relative change of relevant figures of merit under a reflective objective**

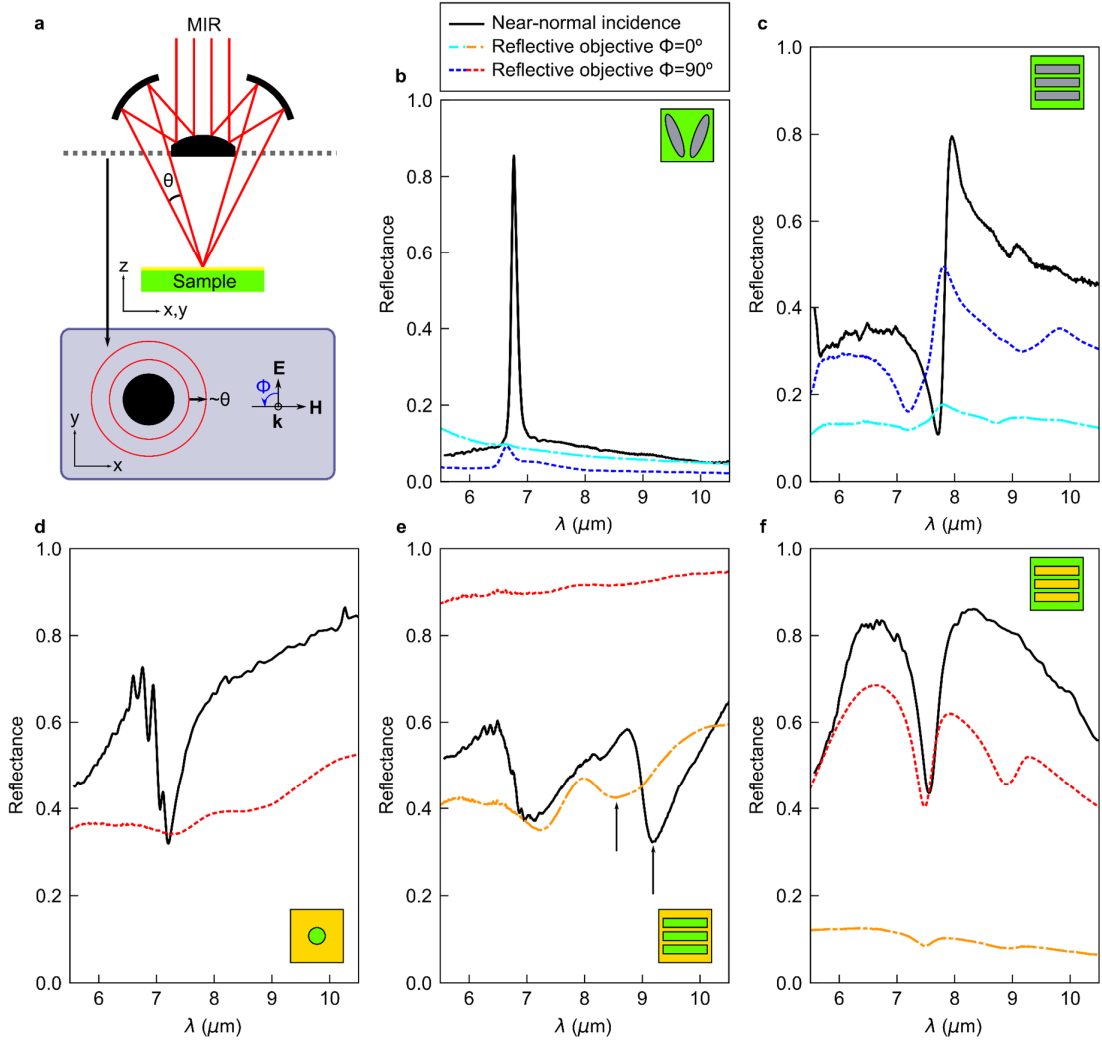

**Figure 7.** Comparison of the superlattice metasurfaces to traditional platforms under a reflective microscope objective. (a) Schematic of the reflective objective used in the Fourier-transform infrared spectrometer indicating the angular stability of the metasurfaces and the definition of the polarization of incident light. (b-f) Comparison of the effect on the metasurface-driven resonances in reflectance by moving from a near-normal incidence illumination polarized to excite the primary resonance (black line) to a reflective objective polarized with $\Phi=0°$ (azure or orange dashed-dotted line) and $90°$ (blue or red dashed line). The metasurfaces shown are the silicon (b) tilted ellipses and (c) silicon superlattice, gold (d) nanohole, (e) inverse and (f) positive superlattice. The q-BIC resonance of the inverse gold superlattice (e) is indicated (black arrows). Schematics of the relevant unit cells are provided in the insets (b-f).

The numerical angular dispersion plots predicted a better performance for the superlattice compared to all-dielectric tilted ellipses or metal nanohole metasurfaces (Figure 4a-e). Using a reflective microscope objective mounted on a Fourier-transform IR (FTIR) spectrometer (**Figure 7a**), we measured the metasurfaces and compared the results with those measured under near-normal incidence (refractive objective, NA=0.15) using mid-IR spectral imaging (Figure 5a). To quantify the performance of the metasurfaces under a switch from a refractive to a reflective objective (NA=0.4), we examine the change in relevant properties or parameters of metasurface-driven resonances, such as the resonance position $\lambda_{res}$, the modulation in reflection $\Delta R$, the Q factor $Q$, and the ratio of coupling constants $\gamma_e/\gamma_i$. Specifically, we introduce a relative change figure of merit (FOM) for the relevant resonance parameters as

$$C = \frac{x_2 - x_1}{x_1}, \qquad (2)$$

where $x_1$ and $x_2$ are the resonance parameters under the refractive and reflective objectives, respectively. With this procedure, the relative change of the resonance position ($C_{\lambda res}$), the Q factor ($C_Q$), the modulation in reflection ($C_{\Delta R}$), and the coupling condition ($C_{\gamma e/\gamma i}$) can be quantified. The closer $C$ is to zero, the better the angular-robust performance of the metasurface is when a switch from a refractive to a reflective objective is performed. To obtain a final performance metric $\bar{C}$ to compare the metasurfaces with each other the relative change FOMs were averaged to a single value. The results are summarized in **Table 1**.

**Table 1.** Comparative summary of the results of the different metasurfaces under the refractive and reflective objectives. The final performance metric $\bar{C}$ for the different metasurfaces was highlighted by appropriately splitting the table.

| Type | Objective | $\lambda_{res}$ (nm) | ~$\Delta R$ | $Q$ | $\gamma_e/\gamma_i$ | $C_{\lambda res}$ | $C_{\Delta R}$ | $C_Q$ | $C_{\gamma e/\gamma i}$ | $\bar{C}$ |
|---|---|---|---|---|---|---|---|---|---|---|
| Tilted ellipses | Refractive | 6764 | 0.75 | 65 | 2.08 | -0.02 | -0.93 | -0.6 | -0.93 | -0.62 |
| | Reflective | 6615 | 0.05 | 26 | 0.14 | | | | | |
| Silicon pos. superlattice | Refractive | 7761 | 0.70 | 31 | 1.27 | -0.02 | -0.57 | -0.39 | -0.73 | -0.43 |
| | Reflective | 7623 | 0.30 | 19 | 0.34 | | | | | |
| Gold nanoholes | Refractive | 7217 | 0.40 | 15 | 0.36 | -1 | -1 | -1 | -1 | -1 |
| | Reflective | - | 0 | 0 | 0 | | | | | |
| Gold inv. superlattice | Refractive | 9050 | 0.30 | 24 | 0.38 | -0.03 | -0.67 | -0.83 | -0.37 | -0.48 |
| | Reflective | 8742 | 0.10 | 4 | 0.24 | | | | | |
| Gold pos. superlattice | Refractive | 7576 | 0.40 | 20 | 0.23 | -0.01 | -0.38 | -0.05 | -0.22 | -0.17 |
| | Reflective | 7474 | 0.25 | 19 | 0.18 | | | | | |

The silicon-based tilted ellipses metasurface showed a strong deterioration of the signals when the switch from near-normal incidence illumination with a refractive objective to the reflective objective was performed (**Figure 7b**) as predicted by the numerical simulations (Figure 4b). The largest sources of deterioration in the resonance parameters came from the reduced modulation in reflection and $\gamma_e/\gamma_i$, each at -0.93. The Q-factor decreased at -0.6 and $\lambda_{res}$ at -0.02.

When the polarization was rotated by 90° no resonance was measured. The switch led to an overall performance $\bar{C}$ of -0.62. This means that by switching the objective from refractive to reflective, the performance of the resonance decreased by 62%. Conversely, 38% of the original resonance parameters remained intact during the increase in the range of incident angles. Consequently, we can quantitatively compare two metasurfaces by evaluating the percentage difference between their performances at maintaining their resonance during the objective switch 1-$\bar{C}$ according to equation (2).

In comparison, the positive silicon superlattice (**Figure 7c**) performed significantly better, with $\bar{C}$ at -0.43 compared to -0.62. Hence, we find that the silicon superlattice performed 50% better than the tilted ellipses. Furthermore, the Q-factor decreased from 31 to 19 at -0.93. Similar to the ellipses, the largest and smallest sources of deterioration of the resonance parameters were $C_{\gamma e/\gamma i}$ and $C_{\lambda res}$ with -0.73 and -0.02, respectively. The silicon superlattice resonance was more robust with respect to maintaining its modulation in reflectance, with $C_{\Delta R}$ at -0.57 compared to -0.93. Additionally, as predicted by the numerical angular dispersion plots (Figure 4a), a smaller second resonance appeared at longer wavelengths under the reflective objective. For *y*-polarized illumination, the resonance was still visible which can be attributed to an imperfect polarizer.

For the gold nanohole metasurface, switching from the refractive to the reflective objective had the biggest impact out of the structures analyzed here (**Figure 7d**). The metasurface-driven resonance that appeared under near-normal incidence vanished under the reflective objective. Hence, $\bar{C}$ was -1. Due to symmetry, the nanohole metasurface is invariant towards rotations of the polarization by 90° and only one polarization had to be measured. In contrast, the resonance of the inverse gold superlattice was measured with the reflective objective with *y*-polarized light (**Figure 7e**), with $\bar{C}$ at -0.48. Compared to the tilted ellipses, the inverse gold superlattice performed 37% better. Similarly to all other metasurfaces studied here, the smallest source of relative change in the resonance parameters of the inverse gold superlattice came from the change in resonance position, at -0.03. Unlike the other metasurfaces, the Q factor contributed to the deteriorating $\bar{C}$ the most with -0.83, where the Q factor decreased from 24 to 4.

The best results were achieved with the positive gold superlattice (**Figure 7f**). Under *x*-polarized illumination, the highly symmetric displacement-mediated q-BIC resonance only marginally deteriorated at -0.17, with the Q-factor decreasing at only -0.05 from 20 to 19. The largest contribution to $\bar{C}$ was $C_{\Delta R}$ at -0.38, although the modulation in reflection with the reflective objective still exceeded 20%. The performance was 118%, 46%, and 59% better than the tilted ellipses, silicon superlattice, and gold inverse superlattice, respectively. A second resonance appeared at longer wavelengths as predicted by the numerical angular dispersion plots (Figure 4c), similar to its silicon counterpart (Figure 7c).

The lower quality resonance of the inverse gold superlattice compared to its positive counterpart can be explained by fabrication limitations due to the negative resist, which required a thinner gold film to allow for a working lift-off and larger gaps between the slots and yielded rounder edges (Figure 5b, SEM images). The quality of the resonance and its angular stability can be improved further by structural optimizations and superior fabrication (see previous discussion in section 2.3). Higher Q factors can easily be achieved for all metasurfaces by decreasing $\alpha$.

## 3. Conclusion

In summary, we have investigated positive and inverse, metal and all-dielectric superlattices and found that they produce displacement-mediated q-BIC resonances that can be described by a displacement asymmetry parameter $\alpha$. Numerical models revealed that an inverse all-dielectric superlattice can produce a displacement-mediated q-BIC resonance under the same polarization as its positive counterpart. Furthermore, the numerical analysis revealed that Babinet's principle applied to the displacement-mediated q-BIC mode for metal superlattices. Numerical investigations of the angular dependence of the q-BIC resonance position successfully led to conclusions on the angular stability of the superlattice variations. The experimental realization of the superlattices confirmed that under near-normal incidence, a displacement-mediated q-BIC resonance emerges for $\alpha \neq 0$. The silicon positive, gold positive and inverse superlattice metasurfaces were fabricated and tested under a reflective microscope objective and compared to two conventional platforms used for sensing: silicon-based tilted ellipses and gold nanohole metasurfaces. We quantified the performance of the metasurfaces by their relative change in their resonance parameters when the range of angles was increased by a switch in microscope objectives. This switch corresponds to an increase in the range of angles and a test of their angular robustness.

The strong performance of the fabricated superlattice metasurfaces compared to rivaling platforms was attributed to better angular robustness predicted by numerical angular dispersion plots. Despite the strong resonances of the silicon tilted ellipses and gold nanohole metasurfaces at near-normal incidence, their resonance parameters deteriorated at -0.62 and -1, respectively. The best performance was achieved with the positive gold superlattice which performed 118%, 46%, and 59% better than the tilted ellipses, silicon superlattice, and gold inverse superlattice, respectively, and was characterized by a highly symmetric line shape. The numerical models showed that a maximum local electric near-field enhancement of ca. 378 can be achieved for $α=0.3$. We anticipate that our experimental results will be useful for the wider research community working with hyper-spectral imaging, FTIR spectroscopy, and laser materials processing. Our results on the angular robustness of the superlattice predict that our metasurfaces can be adopted by researchers working on conformal optical devices, displays, and photonic devices using focused light. In particular, we anticipate that the positive gold superlattice will be used in biospectroscopy and sensing due to the externalized and high electric near-field enhancement and symmetric resonance profile.

## 4. Experimental Section/Methods
*Numerical simulations*

The simulations were performed in CST Studio Suite 2021 using the finite-element frequency-domain Maxwell solver. $CaF_2$ was simulated with a refractive index, n, of 1.35, the surrounding medium as air with n=1, silicon with n=3.48, and gold using the data given in Ref.[49] An impedance-matched open port with a perfectly matched layer introduced linearly polarized light at an azimuthal and polar angle (Figure 1b) through the air boundary into the system. The same port was placed on the opposite side to transmit the transmitted power. The reflectance was recorded by comparing the reflected power to the introduced power. The unit cell (Figure 1c) was defined and then simulated as an infinite periodic array *via* Floquet boundaries. A field monitor recorded the electric near-field distribution. The value of the highest field enhancement of the system was evaluated within the volume of the numerical model. To extract the Q-factor, central wavelength, and $γ_e/γ_i$ of the resonances, the measured spectra were fitted in reflectance (Figure 7) using temporal coupled mode theory.[5]

*Metasurface fabrication*

With $CaF_2$ as the substrate for all fabricated metasurfaces, there were two fabrication protocols, one for the positive and one for the inverse metasurfaces. The metasurfaces were made 100 by 100 μm² in size. The $CaF_2$ windows were cleaned in an acetone bath in an ultrasonic cleaner followed by oxygen plasma cleaning. For the positive metasurfaces, a $CaF_2$ window was spin-coated with positive tone resist (PMMA 950 K A4) baked at 180 °C for 180 s, and with a conducting layer (ESpacer 300Z). For the inverse metasurfaces, a CaF$_2$ window was spin-coated first with an adhesion promoter (Surpass 4000), then with a layer of negative tone resist (ma-N 2403) which was baked at 100 °C for 60 s, and finally with a conducting layer (ESpacer 300Z). The metasurface patterns were created by defining the unit cell and reproducing it in the *x* and *y*-directions. Then, the patterns were written via electron-beam lithography (Raith Eline Plus) with an acceleration voltage of 20 kV for the positive and 30 kV for the inverse metasurfaces, each with an aperture of 20 µm. The exposed resist was developed at room temperature in a 7:3 isopropanol:H$_2$O solution for 50 s for the positive, and in ma-D 525 for 70 s for the inverse metasurfaces. The patterned and developed surface was then coated with a titanium adhesion layer (2 nm at 0.4 Å s$^{-1}$) and a gold film (70 nm for the positive, 30 nm for the inverse metasurfaces at 1 Å s$^{-1}$) using electron-beam evaporation. Finally, an overnight lift-off at 80 °C in Microposit Remover 1165 for the positive, and mr-REM 700 for the inverse metasurfaces concluded the top-down fabrication process. Residual mr-REM 700 was removed in a water bath and by oxygen plasma cleaning. All resists, solutions, and the conducting layer was from micro resist technology GmbH, Germany.

*Mid-IR spectral imaging with a refractive microscope objective*

The near-normal incidence optical responses of the fabricated metasurfaces (Figure 5 and 6) was characterized with a mid-IR spectral imaging microscope Spero from Daylight Solutions Inc., USA, using a low numerical aperture objective (NA=0.15) with a low 4× magnification and a 2 mm$^2$ field of view. The Spero microscope is equipped with four tunable quantum cascade lasers continuously covering the mid-IR spectral range from 948 to 1800 cm$^{-1}$ (ca. 5.56 to 10.5 µm) with a step size of 0.5 cm$^{-1}$. The measurements were conducted in reflection mode and normalized to the reflection signal of a plain gold mirror. The hyperspectral reflectance data corresponding to the metasurfaces were then averaged to reduce noise.

*FTIR imaging with a reflective microscope objective*

A VORTEX 80V FTIR (Bruker) paired with a HYPERION 3000 microscope (Bruker) was used to measure the spectra of the metasurfaces with a reflective microscope objective (Newport) with a NA=0.4 and 15× magnification (Figure 7). The spectra were recorded with a liquid nitrogen-cooled mercury cadmium telluride (LN-MCT) detector in reflection mode. The range of polar angles taken in by the reflective objective was between 12° and 23.6°.

**Supporting Information**

Supporting Information is available from the author.

**Acknowledgements**

The authors thank Thomas Weber for his help with the temporal-coupled mode theory algorithms. This project was funded by the Deutsche Forschungsgemeinschaft (DFG, German Research Foundation) under grant numbers EXC 2089/1–390776260 (Germany's Excellence Strategy) and TI 1063/1 (Emmy Noether Program), the Bavarian program Solar Energies Go Hybrid (SolTech) and the Center for NanoScience (CeNS). Funded by the European Union (projects METANEXT, 101078018 and NEHO, 101046329). Views and opinions expressed are however those of the author(s) only and do not necessarily reflect those of the European Union or the European Research Council Executive Agency. Neither the European Union nor the granting authority can be held responsible for them. S.A.M. additionally acknowledges the Lee-Lucas Chair in Physics and the EPSRC (EP/W017075/1).

Supporting Information

# Metallic and All-Dielectric Metasurfaces Sustaining Displacement-Mediated Bound States in the Continuum

*Luca M. Berger [1], Martin Barkey [1], Stefan A. Maier [2,3,1], Andreas Tittl [1,*]*

[1] *Chair in Hybrid Nanosystems, Nanoinstitute Munich, and Center for NanoScience, Faculty of Physics, Ludwig-Maximilians-University Munich, Königinstraße 10, 80539 München, Germany*
[2] *School of Physics and Astronomy, Monash University, Wellington Rd, Clayton VIC 3800, Australia*
[3] *Department of Physics, Imperial College London, SW7 2AZ London, United Kingdom*
*E-mail: Andreas.Tittl@physik.uni-muenchen.de*

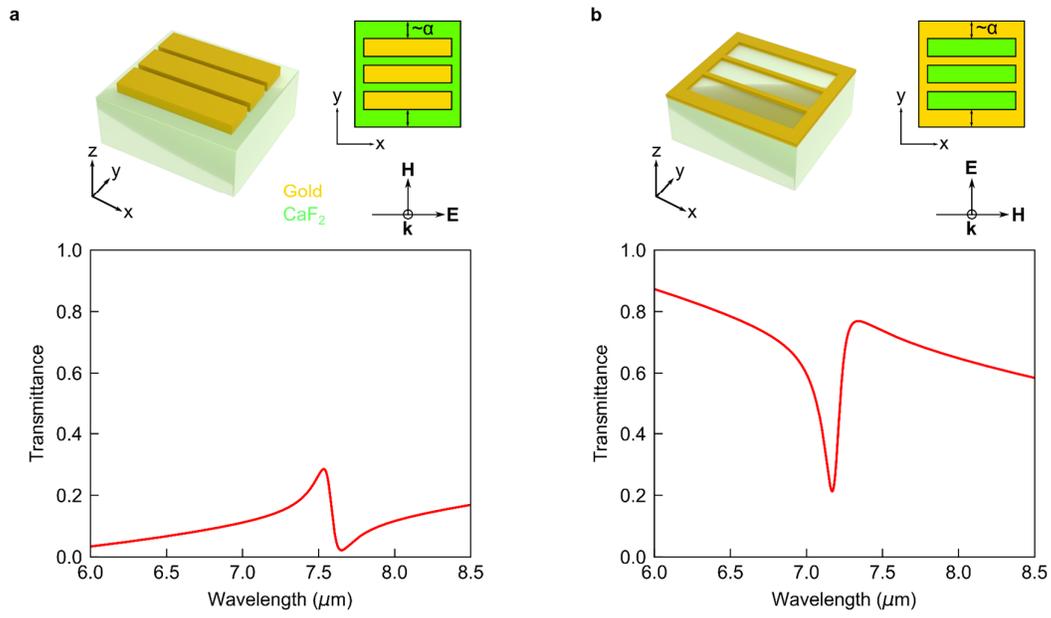

**Figure S1.** The simulated transmission spectra of the gold (a) positive and (b) inverse superlattice for $\alpha$=0.3 at normal incidence with the same parameters as provided in Figure 3. Schematics are included showing the unit cell and polarization of incident light.